\documentclass[10pt,a4paper]{article}
\usepackage[T1]{fontenc}
\usepackage[utf8]{inputenc}
\usepackage{float}
\usepackage{amsmath}
\usepackage{amsfonts}
\usepackage{latexsym}
\usepackage{multicol}
\usepackage[pdftex]{graphicx}
\usepackage{geometry}
\usepackage{verbatim}
\usepackage{xcolor}
\usepackage[normalem]{ulem}

\newgeometry{tmargin=2cm, bmargin=2cm, lmargin=1.5cm, rmargin=1.5cm}

\begin{document}
\pagestyle{empty}
\begin{center}
{\huge\textbf{Molecular beam epitaxy growth of cadmium telluride structures on hexagonal boron nitride}\\}
$\qquad$\\

Adam Krzysztof Szczerba$^1$* , Julia Kucharek$^1$, Jan Pawłowski$^1$, Takashi Taniguchi$^2$, Kenji Watanabe$^3$, \\ and Wojciech Pacuski$^1$**\\
\emph{$^1$Faculty of Physics, University of Warsaw, Pasteura St. 5, 02-093 Warsaw, Poland\\
$^2$Research Center for Materials Nanoarchitectonics, National Institute for Materials Science,  1-1 Namiki, Tsukuba 305-0044, Japan\\
$^3$Research Center for Electronic and Optical Materials, National Institute for Materials Science, 1-1 Namiki, Tsukuba 305-0044, Japan}\\
*email: ak.szczerba@student.uw.edu.pl, **email: Wojciech.Pacuski@fuw.edu.pl\\
$\qquad$\\
\hrule width\textwidth \hfil\break
\end{center}

\begin{multicols*}{2}
\begin{abstract}
\noindent We investigate the feasibility of epitaxial growth of a three-dimensional semiconductor on a two-dimensional substrate. In particular, we report for the first time on molecular beam epitaxy growth of cadmium telluride (CdTe) quantum wells on hexagonal boron nitride (hBN). The presence of the quantum wells is confirmed by photoluminescence measurements conducted at helium temperatures. Growth of quantum wells on two dimensional, almost perfectly flat hBN appears to be very different from growth on bulk substrates, in particular it requires 70-100$^\circ$C lower temperatures.
\end{abstract}

\section{Introduction}
Hexagonal boron nitride (hBN) is a semiconductor with a very high band gap of approximately 5.8 eV$^\text{\cite{Watanabe2004, Yamamoto2023}}$ and ultra-low roughness when it is in form of two-dimensional flakes exfoliated from high quality bulk, such as e.g. bulk grown using high pressure method$^\text{\cite{Taniguchi2007}}$. These properties make hBN an ideal substrate for epitaxial growth, what has been shown for the two-dimensional materials, such as graphene$^\text{\cite{Yang2013}}$  or transition metal dichalcogenides (TMDs) like WS$_2$$^\text{\cite{Okada2014}}$, MoS$_2$$^\text{\cite{Fu2017}}$, MoSe$_2$$^\text{\cite{Poh2018, Pacuski2020}}$, and MoTe$_2$$^\text{\cite{Seredynski2022}}$. In particular, growth on hBN is instrumental for obtaining narrow excitonic lines of TMD monolayers$^\text{\cite{Pacuski2020}}$ without any mechanical postprocessing. Moreover, high bandgap of hBN gives a possibility of using this material as barrier in quantum structures.

The main goal of this work was to verify the effectiveness of growing three-dimensional semiconductors on hBN. Specifically, we decided to grow on hBN CdTe quantum wells (QWs) with (Cd,Mg)Te barrier on top, because optical properties of CdTe/(Cd,Mg)Te quantum wells are extremely sensitive to the quality of the substrate$^\text{\cite{Gaj1994}}$ and to growth conditions. Additionally, in a proposed configuration CdTe layer is in direct contact with hBN, revealing quality of hBN/CdTe interface. For our best knowledge, this is the first report on II-VI semiconductor structure grown on hBN.

\section{Methods}
Growth was performed using molecular beam epitaxy (MBE) in the growth chamber model SVT-35 placed in the University of Warsaw. To grow the samples, low-temperature effusion cells with Cd (7N purity), Mg (6N purity), and Te (7N purity) were used. The evolution of the surface during the growth process was observed with Reflection High-Energy Electron Diffraction (RHEED). With this method, we were able to distinguish between the situation when the hBN surface is covered by deposited material and the situation when efficient desorption leads to clean hBN despite exposition to molecular fluxes. Additionally, we observed that the exposition of the substrate on the electron beam affects slightly growth conditions, as described in section Results and Discussion. After growth, the sample surface was imaged using optical microscopy and Atomic Force Microscopy (AFM). The presence of the quantum wells was verified through photoluminescence measurements conducted at a temperature of 10 K under a microscope objective with a laser spot of about 1 micrometer diameter. A laser with a wavelength of 445 nm was used to excite the samples. This wavelength corresponds to a photon energy of 2.8 eV, so it is sufficient to excite valence band electrons to the conduction band of CdTe, considering that the band gap of CdTe at 10 K is  1.6 eV$^\text{\cite{Fonthal2000}}$. Excitation power was relatively low (300 $\mu$W) to avoid structural influence of the laser beam on the studied structure.

\section{Design of the quantum well structure on hBN}
\begin{figure}[H]
\centering
\includegraphics[width=0.5\textwidth]{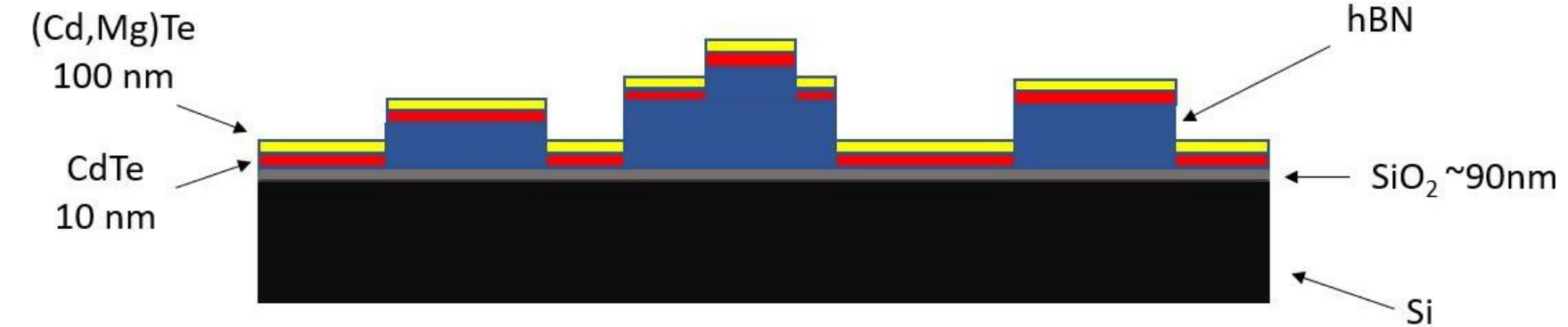}\\
\caption{Scheme of CdTe/(Cd,Mg)Te QW sample grown by MBE on hBN flakes exfoliated on Si (100) wafer covered with 90 nm of SiO$_2$. The thickness of CdTe is 10 nm, the thickness of (Cd,Mg)Te is 100 nm, the thickness of hBN flakes varies between a few nm and a few hundreds of nm, typically about 100 nm. }
$\qquad$\\
\hrule width0.5\textwidth \hfil\break
\label{scheme}
\end{figure}

Figure \ref{scheme} illustrates the design of the samples investigated in this work. The substrates were prepared by exfoliating hBN flakes onto a semi-insulating, 10 mm large Si (100) wafer with 90 nm of SiO$_2$. The bulk hBN used during the experiment was a high-quality material grown in the laboratory of K. Watanabe and T. Taniguchi$^\text{\cite{Taniguchi2007}}$. The exfoliated hBN flakes had an average thickness of approximately 100 nm and typical size of tens of micrometers. On such substrate, a layered structure was grown through MBE, involving the growth of nominally 10 nm of CdTe as well as 100 nm of a barrier material, primarily (Cd,Mg)Te with about 10 percent of Mg. Since hBN flakes were covering SiO$_2$ only partially, CdTe structures have been deposited at the same time on both hBN and SiO$_2$.

\begin{figure}[H]
\centering
\includegraphics[width=0.35\textwidth]{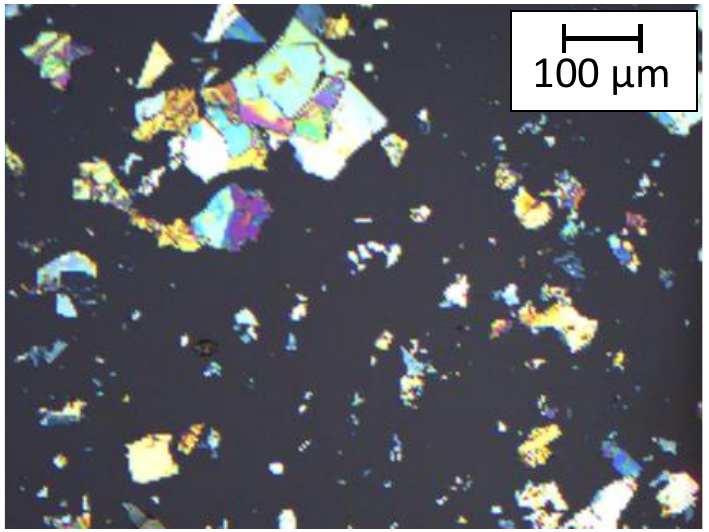}\\
\caption{Example optical image of Si (100) substrate covered with 90 nm of SiO$_2$ and exfoliated hBN flakes with various thickness and lateral size. Such substrates were further used for epitaxial growth of II-VI structures.}
$\qquad$\\
\hrule width0.5\textwidth \hfil\break
\label{hBN_wide}
\end{figure}

Optical image of part of the substrate’s surface is presented in Fig. \ref{hBN_wide}. Many hBN flakes with different shapes, sizes and colors are visible. The color of the flake corresponds to its thickness, which means that each flake has a different height (from a few to hundred nanometers). Boron nitride flakes have also a huge range of lateral size, up to hundreds of micrometers. 

\begin{figure}[H]
\centering
\includegraphics[width=0.4\textwidth]{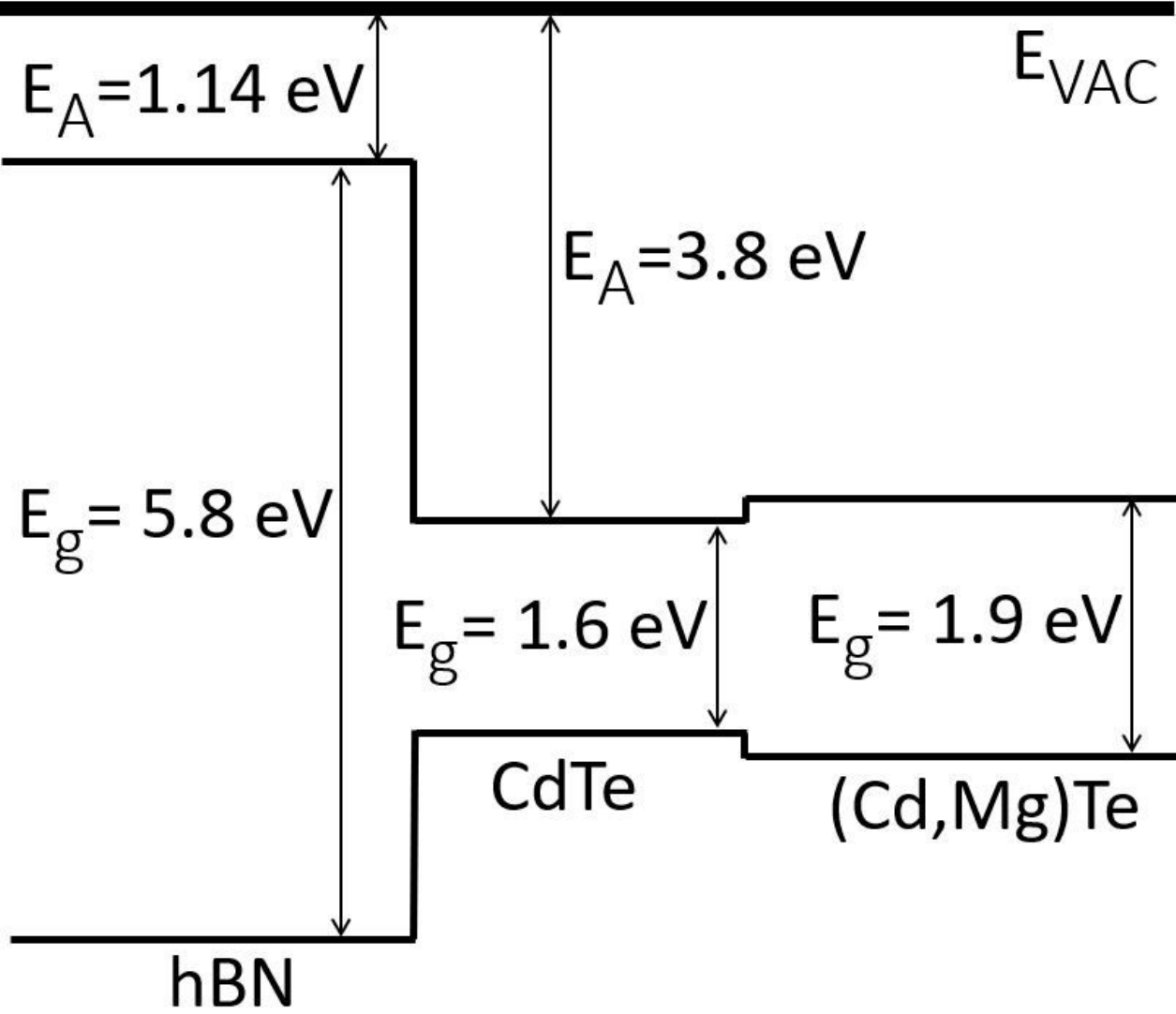}\\
\caption{Scheme of band structure of CdTe/(Cd,Mg)Te QW grown on hBN. $E_{VAC}$ is a vacuum level, $E_A$ is electron affinity and $E_g$ is energy gap of used materials.}
$\qquad$\\
\hrule width0.5\textwidth \hfil\break
\label{band_scheme}
\end{figure}

Band structure of CdTe/(Cd,Mg)Te grown on hBN is presented in Fig. \ref{band_scheme}. Both hBN/CdTe and CdTe/(Cd,Mg)Te interfaces are I type heterojunction, which means the quantum well made of these materials also is I type and its optical properties are promising. Valence band offset $\alpha$ was calculated by dividing the valence band energy level $E_v$ difference of both materials X and Y by band gap energy $E_g$ difference$^\text{\cite{Gaj1994}}$:
\begin{displaymath}
\alpha_{X/Y}=\frac{E_v(X)-E_v(Y)}{E_g(Y)-E_g(X)}
\end{displaymath}

The valence band offset of hBN and CdTe was estimated based on electron affinity for these materials$^\text{\cite{Knobloch2021, Mamta2022}}$. We estimate a value of valence band offset of CdTe and hBN to be approximately $\alpha_{\text{CdTe/hBN}}=0.37$. Valence band offset of CdTe and (Cd,Mg)Te is $\alpha_{\text{CdTe/(Cd,Mg)Te}}=0.45$$^\text{\cite{Wojtowicz1998}}$.

\section{Results and Discussion}
The technological experiment  was started by calibration of growth rates (approximately 0.1 nm/s) using in-situ optical reflectivity during growth of standard of CdTe and (Cd,Mg)Te layers on GaAs (100) substrate$^\text{\cite{Pacuski2017}}$. Then we started growth of similar structures on hBN/SiO$_2$/Si substrates, at the same temperature, which was equal to 320°C. However, neither RHEED observation nor post growth optical and AFM imaging reveled presence of deposited material. We concluded, that at such a temperature sticking coefficient is zero for perfectly flat surfaces. There was some material grown on SiO$_2$ on part of the sample, but even on hBN flakes surrounded by grown material on SiO$_2$, there was no material observed. Consequently, we substantially decreased substrate temperature during the growth of the next samples.

\begin{figure}[h!]
\centering
\includegraphics[width=1\textwidth]{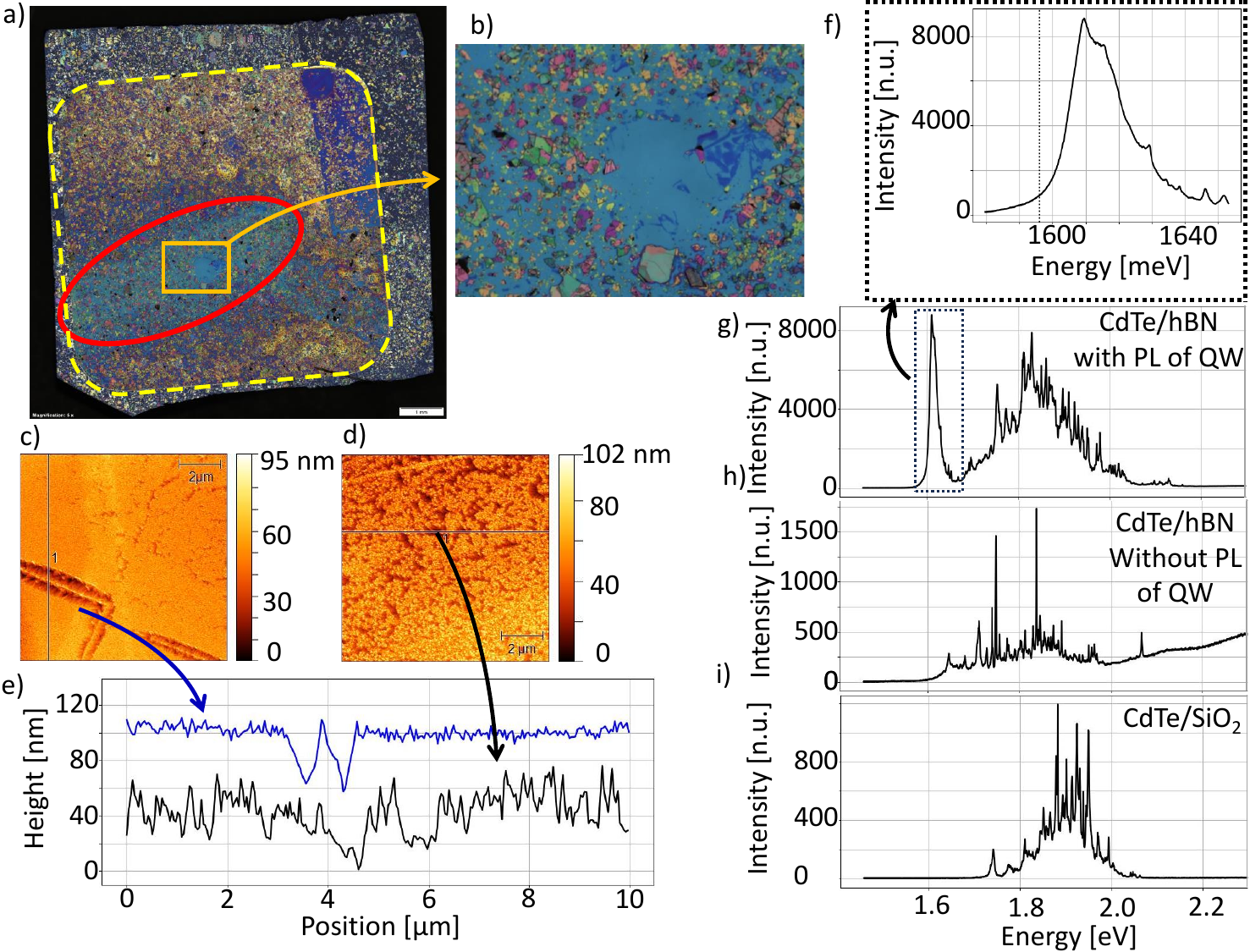}\\
\caption{A sample with CdTe QW deposited at 250°C on hBN flakes exfoliated on Si, all covered with (Cd,Mg)Te barrier a) Optical image of whole 8.5 mm x 9 mm sample. The sample is covered by exfoliated hBN flakes like in Fig. \ref{hBN_wide}. The growth has occurred only in area marked by a dashed square in the middle of the sample, and area outside the square was not exposed to the molecular fluxes. Area influenced by high-energy electrons from RHEED gun was marked by ellipsoid. b) Magnification of part of the sample marked in Fig. \ref{panel}a. In the middle of the image area without exfoliated hBN is visible. This area is surrounded by many hBN flakes of different sizes and colors. c) AFM scan of material grown on hBN in area influenced by RHEED. d) AFM scan of material grown on hBN in area without influence of RHEED. e) Comparison of roughness of material grown on hBN in area influenced by RHEED (blue line, RMS=3.426 nm) and in area without such influence (black line, RMS=6.1 nm)  f) Magnification of PL spectrum presented in Fig. \ref{panel}g with CdTe QW signal. g) Broad range PL spectrum of the structure measured at 10 K. A strong peak in characteristic energy close to 1610 meV is identified as related to CdTe QW. Multiple peaks appeared in a wide range of energies were associated with PL signal of the barrier. h) PL spectrum of structure in areas where the presence of CdTe QW is not evident. Multiple peaks appeared in a wide range of energies were associated with PL signal of the barrier. i) Typical PL spectrum of the structure grown in the same process but on SiO$_2$. Multiple peaks appeared in a wide range of energies were associated with PL signal of the barrier. }
$\qquad$\\
\hrule width\textwidth \hfil\break
\label{panel}
\end{figure}

The sample presented in Fig. \ref{panel} was grown at a temperature of 250°C, which is approximately 70°C lower than the growth temperature typically used for growth of CdTe QWs on bulk substrates. Optical image (Fig. \ref{panel}a,b) of the sample reveals a high number of hBN flakes in the whole area of the sample. Almost whole sample is covered by deposited layer, except edges which were not exposed to molecular fluxes. Observed blue-violet color of deposited layer results from optical interferences. Subtle changes in the color is a consequence of small differences in layer thickness and resulting optical interferences. They reveal in particular an area marked by red ellipsoid, which was exposed to electron beam related to RHEED measurements during growth without substrate rotation. AFM was used to scan the surface of the material grown on hBN on different regions on the sample. In particular AFM reveals a difference between areas which are effected by electron beam (Fig. \ref{panel}c), and other areas (Fig. \ref{panel}d).
\end{multicols*}

\begin{multicols*}{2}

\noindent (Cd,Mg)Te grown on hBN appears to be more compact in the region influenced by electrons. Cross-sections of AFM scans in Fig. \ref{panel}c,d are presented in Fig. \ref{panel}e. In area influenced by RHEED typical height difference is approximately 10 nm, however 40 nm deep valeys, caused by atomic steps on the substrate, are visible. Cross-section of area without electron beam influence presents huge height differences in short distance on the sample. The largest difference in this graph is approximately 80 nm in range of 1 $\mu$m, which is 80\% of nominal thickness of (Cd,Mg)Te barrier. Representative photoluminescence spectra measured in various spots of the sample are shown in the Fig. \ref{panel}f,g,h for the structure grown on hBN and in the Fig. \ref{panel}i for the structure grown on SiO$_2$.

Typical PL signal of CdTe QW grown on hBN is shown in Fig. \ref{panel}g, with an intense peak located at 1610 meV, so in spectral position typical for CdTe QWs grown on bulk substrates, only slightly blue-shifted from position known for bulk, 1596 meV$^\text{\cite{Gaj1994}}$, what is visible in Fig. \ref{panel}f. Interestingly, such PL spectra are observed mainly in the area of the sample, which was affected by RHEED observations. Fig. \ref{panel}h show areas where PL of CdTe QW on hBN is less evident, it is at higher energy (position 1645 meV), intensity is weaker and the peak is merged with an ensemble of sharp lines.  Emission energy of the QW strongly depends on the thickness of the quantum well, but can be also increased by interdiffusion of Mg from (Cd,Mg)Te barrier. In our case, both reasons can be responsible for the observation of the first peak at higher energy than usual, as it is in Fig. \ref{panel}h. 

\begin{figure}[H]
\centering
\includegraphics[width=0.5\textwidth]{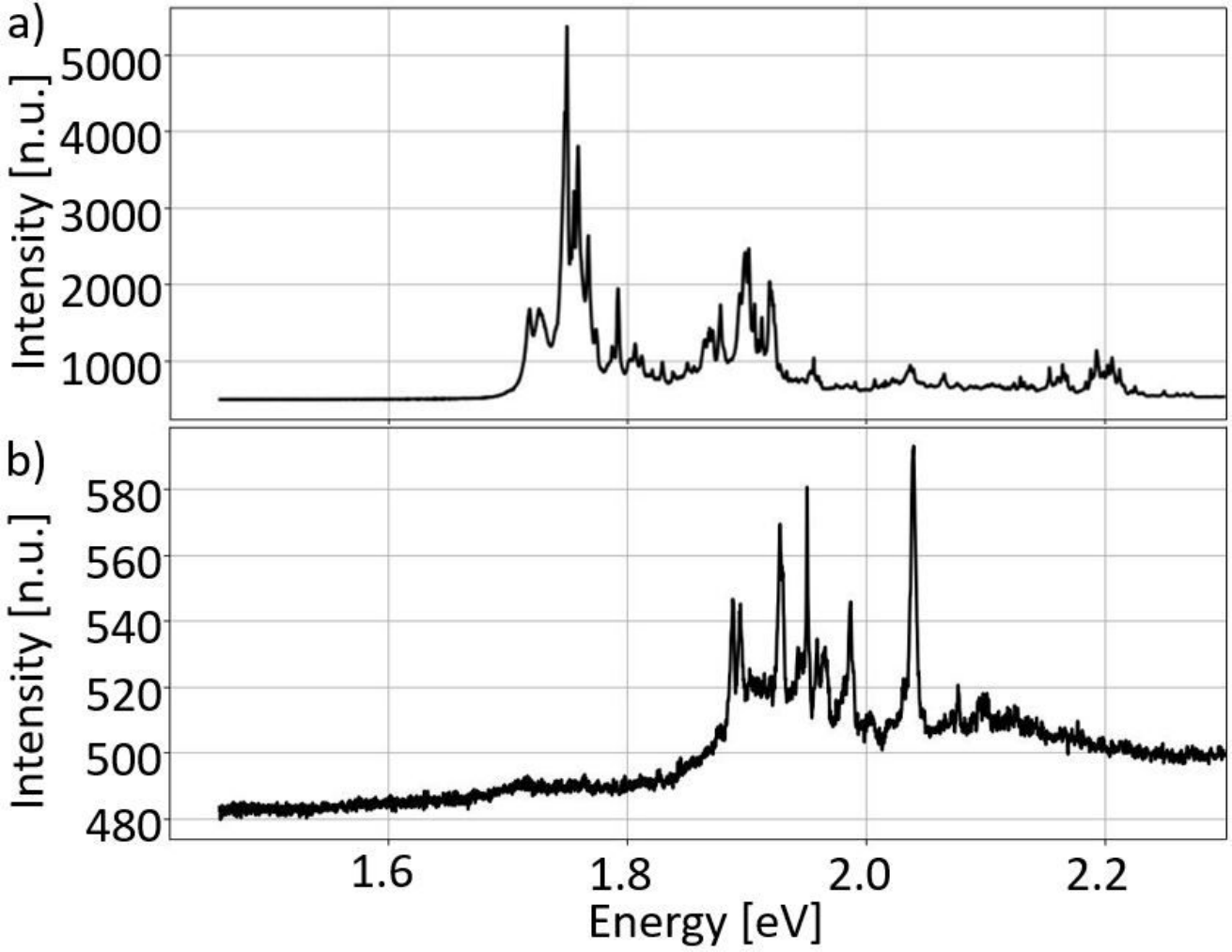}\\
\caption{PL spectra of a reference (Cd,Mg)Te layer, without a QW, measured at temperature 10 K. Multiple peaks appeared in a wide range of energies were associated with PL signal of the barrier. a) Typical PL spectrum of 100 nm (Cd,Mg)Te grown on hBN. b) Typical PL spectrum of 100 nm (Cd,Mg)Te grown on SiO$_2$.}
$\qquad$\\
\hrule width0.5\textwidth \hfil\break
\label{barrierPL}
\end{figure}

In the Fig. \ref{panel}i for the structure grown on SiO$_2$ there is no clear peak close to expected emission energy of CdTe QWs, only ensemble of sharp lines is observed. Similar PL spectrum containing many sharp lines in the range 1.7-2.05 eV and observed for all areas of the sample. In order to explain the origin of multiple lines observed in a wide range, we have grown and studied a reference sample where only (Cd,Mg)Te barrier was deposited, without CdTe QW. PL of such reference sample is shown in Fig. \ref{barrierPL}a for (Cd,Mg)Te deposited on hBN, and in Fig. \ref{barrierPL}b for (Cd,Mg)Te deposited on SiO$_2$. In both cases, multiple sharp lines in a wide range are observed. This indicates, that such sharp lines are not related to CdTe QW, they are related just to (Cd,Mg)Te barrier. Lines of (Cd,Mg)Te observed at various energies indicate a high structural and compositional disorder of this material. This is consistent with results obtained for bulk (Cd,Mg)Te and epitaxial (Cd,Mg)Te on 3D substrates, where the tendency to separation of various phases is observed$^\text{\cite{Yu2019, Leblanc2017}}$. Based on the emission energy in the wide range between 1.7 eV and 2.2 eV (e.g. Fig. \ref{barrierPL}a), composition in various grains corresponds to Mg concentration between 5\% and 30\%.

\begin{figure}[H]
\centering
\includegraphics[width=0.5\textwidth]{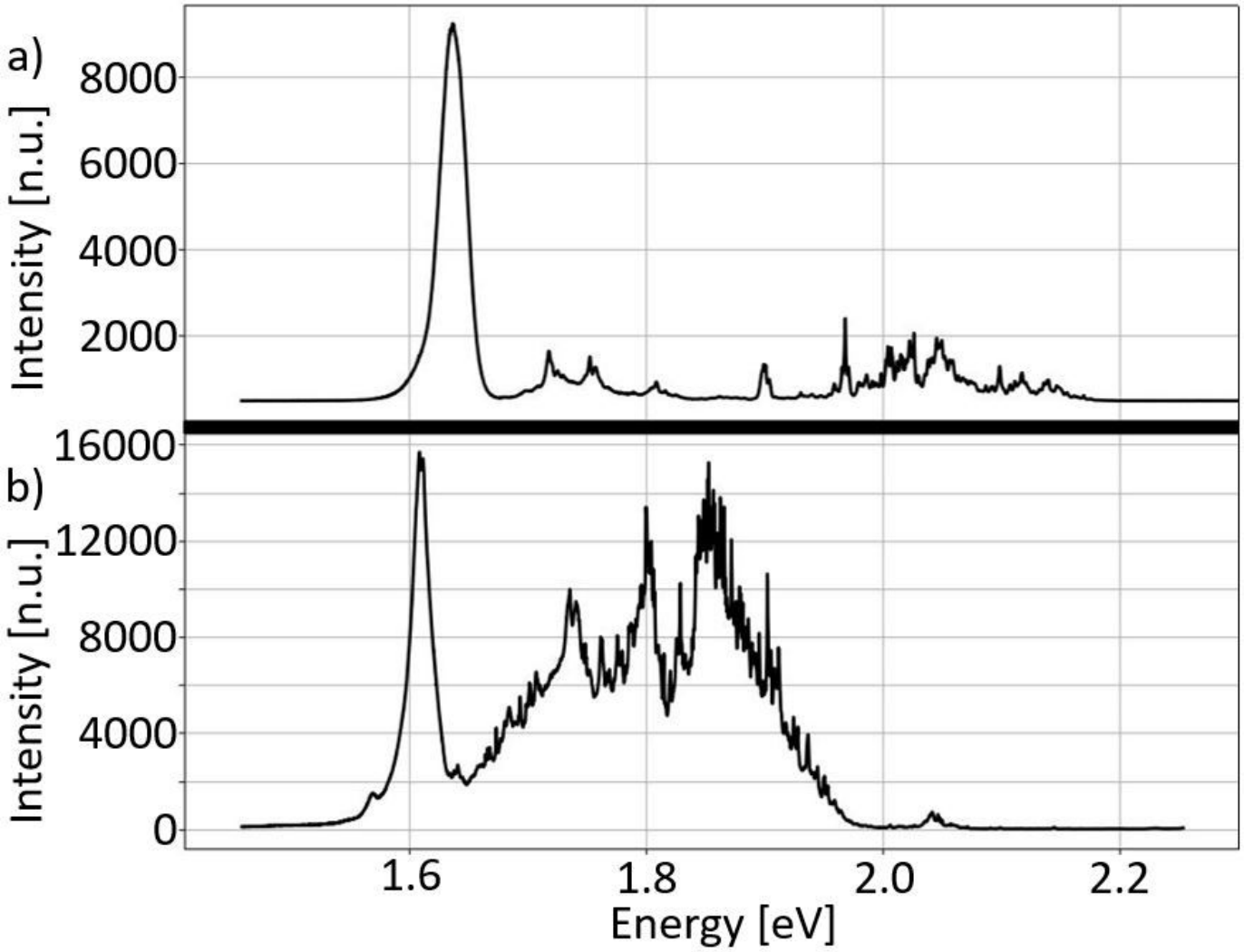}\\
\caption{a) Typical PL spectra of 10 nm CdTe and 100 nm (Cd,Mg)Te grown on hBN in temperature 100$^\circ$C lower than the growth temperature on three-dimensional material. Multiple peaks appeared in a wide range of energies were associated with PL signal of the barrier. b) Typical PL spectra of 10 nm CdTe and 100 nm (Cd,Mg)Te grown on hBN annealed before the growth. Multiple peaks appeared in a wide range of energies were associated with PL signal of the barrier.
}
$\qquad$\\
\hrule width0.5\textwidth \hfil\break
\label{220C_and_annealed}
\end{figure}

Throughout this experiment, multiple samples were grown to identify the best conditions for the growth. As a result, it was found that the optimal substrate temperature during growth was 220$^\circ$C, approximately 100°Clower than the growth temperature on three-dimensional material. Typical PL spectrum of CdTe QW grown on hBN in this conditions is illustrate in Fig. \ref{220C_and_annealed}a. Observed peak of QW is much stronger than in the case of the first grown sample presented in Fig. 2a, grown at 250°C.

In order to understand the role of electron beam in the formation of QWs, we performed experiment with high temperature (800$^\circ$C) annealing of the substrate before growth of QWs. In such structures, there was no traces of electron beam any more, and QWs were observed in the whole area where CdTe/(Cd,Mg)Te was deposited on hBN (Fig. \ref{220C_and_annealed}b). Therefore electron beam acts in similar way as degassing at high temperature. Another conclusion is that degassing substrate at about 200$^\circ$C just before the growth is not enough to clean the surface properly.

\begin{figure}[H]
\centering
\includegraphics[width=0.9\textwidth]{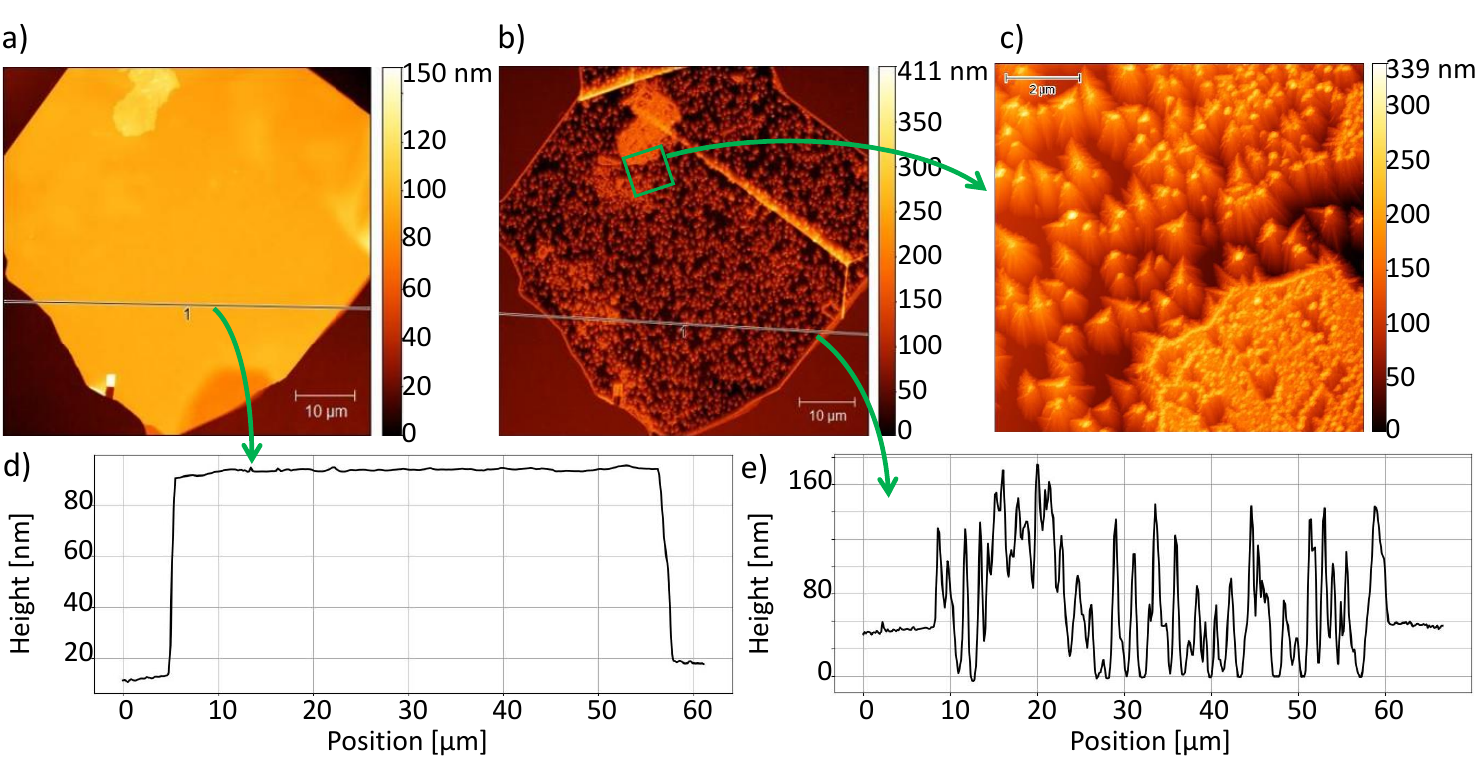}\\
\caption{AFM scans and cross sections of the hBN flake before and after the growth of 10 nm CdTe and 100 nm (Cd,Mg)Te performed in 220$^\circ$C on the substrate annealed in 800$^\circ$C before the growth process. a) AFM scan of hBN flake before the growth. b) AFM scan of hBN flake after the growth. The evolution of the flake’s substrate in comparison to the image shown in Fig. \ref{UW2148_hBN_comp}a is visible, which means that growth has occurred. c) AFM scan of part of the surface of material grown on hBN (marked by a square in Fig. \ref{UW2148_hBN_comp}b). Many structures similar to triangle-based pyramids are visible, indicating the crystal orientation (111) of the growth. d) Cross section of AFM scan of hBN flake before the growth performed along the line marked in Fig. \ref{UW2148_hBN_comp}a. The flake has a height of 80 nm, and the RMS calculated on hBN along the line was approximately 58.49 pm. e) Cross section of AFM scan of material grown on hBN performed along the line marked in Fig. \ref{UW2148_hBN_comp}b. The height level of material grown on SiO$_2$ and on hBN is similar, which indicates different conditions of growth on both surfaces. RMS of the material grown on hBN calculated along the line is approximately 12.06 nm.}
$\qquad$\\
\hrule \hfil\break
\label{UW2148_hBN_comp}
\end{figure}

AFM scans of the sample of 10 nm CdTe and 100 nm (Cd,Mg)Te grown in the best conditions (growth performed in 220$^\circ$C on the substrate annealed in 800$^\circ$C) and scan of the same flake before the growth were shown in Fig. \ref{UW2148_hBN_comp}a,b,c. The evolution of the sample’s surface is clearly visible. The cross-section of the hBN flake before and after the growth is presented in Fig. \ref{UW2148_hBN_comp}d,e.

\end{multicols*}

\begin{multicols*}{2}

On the cross-section presented in Fig. \ref{UW2148_hBN_comp}d, the thickness of the hBN flake before growth was determined to be approximately 80 nm. The RMS of hBN flakes along the cross-section line is 58.49 pm. After the growth on the hBN structures similar to pyramids are visible and the RMS on the sample along the cross-section line is approximately 12.06 nm. Material grown on SiO$_2$ is more dense (RMS=7.94) than the material on hBN. Furthermore, the height level of material grown on SiO$_2$ is similar to the height level of material on hBN, which shows how the growth conditions in both areas are different.

Many pyramid structures visible in Fig. \ref{UW2148_hBN_comp}c have a triangular base. This observation indicates that trigonal symmetry of the substrate resulted in the growth of CdTe and (Cd,Mg)Te with the same symmetry, therefore in (111) crystallographic direction.

\begin{figure}[H]
\centering
\includegraphics[width=0.35\textwidth]{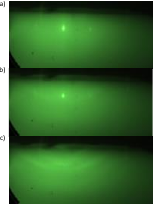}\\
\caption{RHEED images obtained during the growth at 220$^\circ$C on the substrate annealed in 800$^\circ$C on various stages. a) the hBN substrate before growth. b) The hBN with nominally 10 nm of CdTe. c) Completed CdTe/(Cd,Mg)Te structre. Rings indicate the polycrystalline structure.}
$\qquad$\\
\hrule width0.5\textwidth \hfil\break
\label{RHEED}
\end{figure}

During the growth of the samples, the changes on the surface were observed with RHEED signal. Example images of such measurements are presented in Fig. \ref{RHEED}, for the growth performed at 220$^\circ$C on the substrate annealed at 800$^\circ$C before the growth process. Well defined diffraction pattern on the image Fig. \ref{RHEED}a originates from the hBN surface, while scattering from SiO$_2$ contributes to the background. After the growth of the nominally 10 nm thick CdTe layer, the signal of hBN is weakening and the deposited material appears as a delicate ring. After the growth of the whole sample, the hBN signal completely disappears and many rings are visible. This indicates that the whole hBN was covered by polycrystalline material.

About 10 nm thick layer of CdTe should form a structure, in which quantum effects are significant vertically and neglectable laterally. Therefore such structure can be considered as a quantum well. Moreover, the observed emission energy, about 1610 meV, agrees well with the characteristic emission energy of CdTe QWs grown on other substrates$^\text{\cite{Wojtowicz1998}}$. 

\section{Conclusion}
Several CdTe QWs on hBN samples were grown using MBE with lowering of substrate temperature comparing to growth on bulk substrates. The first QW was detected on the sample grown in temperature approximately 70°C lower than the typical growth temperature typically used for bulk materials. Typical photoluminescence signal of CdTe QW deposited directly on hBN is observed close to 1610 meV, similarly as for well known QWs CdTe/(Cd,Mg)Te. Furthermore, the surface's structure of (Cd,Mg)Te grown on hBN was analysed through AFM scans. The properties of the barrier material were found to be connected with the broad PL spectra presented in Fig. \ref{barrierPL}a,b. This spectra appears in all PL signals of the samples with (Cd,Mg)Te barrier, as shown in Fig. \ref{panel}f,g,h and Fig. \ref{220C_and_annealed}a,b,c.

The optimal temperature of the substrate was found to be 220$^\circ$C, what is approximately 100$^\circ$C lower than the growth temperature on three-dimentional material. In this growth temperature PL signal of CdTe QW (Fig. \ref{220C_and_annealed}a) was observed over the majority of the hBN. For effective growth of CdTe QWs on hBN, the substrate should be influenced by high-energy electron beam or pre-heated to 800$^\circ$C before the growth.

Studied hBN/CdTe/(Cd,Mg)Te heterostructure appears to be a new, high optical quality type I QW that benefits from the ultra-high flatness of 2D barrier. This opens an exciting possibility of redesigning various QWs systems by replacing the bottom barrier with hBN.

\section{Acknowledgments}
We acknowledge the financial support from National Science Centre Poland, project no. 2021/41/B/ST3/04183. K.W. and T.T. acknowledge support from the JSPS KAKENHI (Grant Numbers 21H05233 and 23H02052) and World Premier International Research Center Initiative (WPI), MEXT, Japan

\end{multicols*}


\begin{thebibliography}{30}

\bibitem{Watanabe2004} Watanabe, K.; Taniguchi, T.; \& Kanda, H.; Direct-bandgap properties and evidence for ultraviolet lasing of hexagonal boron nitride single crystal. \emph{Nature Mater.}  2004, 3, 404–409, DOI: 10.1038/nmat1134.

\bibitem{Yamamoto2023} Yamamoto, M.; Murata, H.; Miyata, N.; Takashima, H.; Nagao, M.; Mimura, H.; Neo, Y.; Murakami, K.; Low-temperature direct synthesis of multilayered h-BN without catalysts by inductively coupled plasma-enhanced chemical vapor deposition. \emph{ACS Omega} 2023, 8 (6), 5497-5505, DOI: 10.1021/acsomega.2c06757.

\bibitem{Taniguchi2007} Taniguchi, T.; \& Watanabe, K.; Synthesis of high-purity boron nitride single crystals under high pressure by using Ba–Bn solvent. \emph{J. Cryst. Growth} 2007, 303, 525–529, DOI: 10.1016/j.jcrysgro.2006.12.061.

\bibitem{Yang2013} Yang, W.; Chen, G.; Shi, Z.; Liu, C.-C.; Zhang, L.; Xie, G.; Cheng, M.; Wang, D.; Yang, R.; Shi, D. Epitaxial Growth of Single-domain Graphene on Hexagonal Boron Nitride. \emph{Nat. Mater.} 2013, 12, 792– 797,  DOI: 10.1038/nmat3695


\bibitem{Okada2014}  Okada, M.; Sawazaki, T.; Watanabe, K.; Taniguchi, T.; Hibino, H.; Shinohara, H.; Kitaura, R. Direct chemical vapor deposition
growth of WS2 atomic layers on hexagonal boron nitride. \emph{ACS Nano} 2014, 8, 8273--8277, DOI: 10.1021/nn503093k

\bibitem{Fu2017} Fu, D.; Zhao, X.; Zhang, Y.-Y.; Li, L.; Xu, H.; Jang, A.-R.; Yoon, S. I.; Song, P.; Poh, S. M.; Ren, T.; Ding, Z.; Fu, W.; Shin, T. J.; Shin, H. S.; Pantelides, S. T.; Zhou, W.; \& Loh, K. P.; Molecular Beam Epitaxy of Highly Crystalline Monolayer Molybdenum Disulfide on Hexagonal Boron Nitride \emph{J. Am. Chem. Soc.} 2017, 139, 27, 9392--9400, DOI: 10.1021/jacs.7b05131.

\bibitem{Poh2018} Poh, S. M.; Zhao, X.; Tan, S. J. R.; Fu, D.; Fei, W.; Chu, L.; Jiadong, D.; Zhou, W.; Pennycook, S. J.; Castro Neto, A. H.; \& Loh, K. P.; Molecular beam epitaxy of highly crystalline MoSe2 on hexagonal boron nitride, \emph{ACS Nano} 2018, 12, 7562, DOI: 10.1021/acsnano.8b04037.

\bibitem{Pacuski2020} Pacuski, W.; Grzeszczyk, M.; Nogajewski, K.; Bogucki, A.; Oreszczuk, K.; Kucharek, J.; Połczyńska, K.E.; Seredyński, B.; Rodek, A.; Bożek, R.; Taniguchi, T.; Watanabe, K.; Kret, S.; Sadowski, J.; Kazimierczuk, T.; Potemski, M.; \& Kossacki, P.; Narrow Excitonic Lines and Large-Scale Homogeneity of Transition-Metal Dichalcogenide Monolayers Grown by Molecular Beam Epitaxy on Hexagonal Boron Nitride. \emph{Nano Letters} 2020, 20, 5, 3058-3066, DOI: 10.1021/acs.nanolett.9b04998

\bibitem{Seredynski2022} Seredyński, B.; Bożek, R.; Suffczyński, J.; Piwowar, J.; Sadowski, J.; \& Pacuski, W.; Molecular beam epitaxy growth of MoTe$_2$ on hexagonal boron nitride. \emph{JCG} 2022, 596, 126806, DOI: 10.1016/j.jcrysgro.2022.126806.

\bibitem{Gaj1994} Gaj, J.A.; Grieshaber, W.; Bodin-Deshayes, C.; \& Cibert, J.; Magneto-optical study of interface mixing in the CdTe-(Cd,Mn)Te system. \emph{Phys. Rev. B} 1994, 50, 8, 5512-5527, DOI: 10.1103/PhysRevB.50.5512

\bibitem{Fonthal2000} Fonthal, G.; Tirado-Mejia, L.; Marin-Hurtado, J. I.; Ariza-Calderon, H.; \& Mendoza-Alvarez, J. G.; Temperature dependence of the band gap energy of crystalline CdTe. \emph{J. Phys. Chem. Solids} 2000, 61, 579–583, DOI: 10.1016/S0022-3697(99)00254-1.

\bibitem{Knobloch2021} Knobloch, T.; Illarionov, Y.Y.; Ducry, F.; Schleich, C.; Wachter, S.; Watanabe, K.; Taniguchi, T.; Mueller, T.; Waltl, M.; Lanza, M.; Vexler, M. I.; Luisier, M.; \& Grasser, T.; The performance limits of hexagonal boron nitride as an insulator for scaled CMOS devices based on two-dimensional materials. \emph{Nat. Electron.} 2021, 4, 98–108, DOI: 10.1038/s41928-020-00529-x

\bibitem{Mamta2022}Mamta; Maurya, K.K.; \& Singh, V.N.; Comparison of Various Thin-Film-Based Absorber Materials: A Viable Approach for Next-Generation Solar Cells. \emph{Coatings} 2022, 12, 405, DOI: 10.3390/coatings12030405

\bibitem{Wojtowicz1998} Wojtowicz, T.; Kutrowski, M.; Karczewski, G.; \& Kossut, J.; Graded quantum well structures made of diluted magnetic semiconductors. \emph{Acta Physica Polonica A} 1998, 94(2), 199-217, DOI: 10.12693/APhysPolA.94.199

\bibitem{Pacuski2017} Pacuski, W.; Rousset, J.-G.; Delmonte, V.; Jakubczyk, T.; Sobczak, K.; Borysiuk, J.; Sawicki, K.; Janik, E.; \& Kasprzak, J. Antireflective photonic structure for coherent nonlinear spectroscopy of single magnetic quantum dots. \emph{Cryst. Growth Des.} 2017, 17, 2987- 2992. DOI: 10.1021/acs.cgd.6b01596


\bibitem{Yu2019} Yu, P.; Jiang, B.; Chen, Y.; Zheng, J.; \& Luan, L.; Study on In-Doped CdMgTe Crystals Grown by a Modified Vertical Bridgman Method Using the ACRT Technique. \emph{Materials} 2019, 12(24), 4236, DOI: 10.3390/ma12244236

\bibitem{Leblanc2017} LeBlanc, E. G.; Edirisooriya, M.; Ogedengbe, O. S.; Noriega, O. C.; Jayathilaka, P. A. R. D.; Rab, S.; Swartz, C. H.; Diercks, D. R.; Burton, G. L.; Gorman, B. P.; Wang, A.; Barnes, T. M.; \& Myers, T. H.; Determining and Controlling the Magnesium Composition in CdTe/CdMgTe Heterostructures. \emph{J. El. Mat.} 2017, 46, 5379–5385, DOI: 10.1007/s11664-017-5589-3


\end{thebibliography}
\end{document}